\documentclass[acmsmall]{acmart}
\usepackage{siunitx}  

\newcommand{\e}{\mathrm{e}}

\setcopyright{acmcopyright}
\copyrightyear{0000}
\acmYear{0000}
\acmDOI{XXXXXXX.XXXXXXX}

\acmJournal{TECS}
\acmVolume{00}
\acmNumber{0}
\acmArticle{000}
\acmMonth{0}

\acmSubmissionID{64}

\begin{document}

\title{%
Modular DFR: Digital Delayed Feedback Reservoir Model for Enhancing Design Flexibility
}

\author{Sosei Ikeda}
\affiliation{%
  \institution{Kyoto University}
  \city{Kyoto}
  \country{Japan}}
\email{sikeda@easter.kuee.kyoto-u.ac.jp}

\author{Hiromitsu Awano}
\affiliation{%
  \institution{Kyoto University}
  \city{Kyoto}
  \country{Japan}} 
\email{awano@i.kyoto-u.ac.jp}

\author{Takashi Sato}
\affiliation{%
  \institution{Kyoto University}
  \city{Kyoto}
  \country{Japan}}
\email{takashi@i.kyoto-u.ac.jp}

\renewcommand{\shortauthors}{Ikeda et al.}

\begin{abstract}

A delayed feedback reservoir (DFR) is a type of reservoir computing system well-suited for hardware
implementations owing to its simple structure.
Most existing DFR implementations use analog circuits that require both
digital-to-analog and analog-to-digital converters for interfacing.
However, digital DFRs
emulate analog nonlinear components in the digital domain, resulting in a lack of design flexibility and higher power consumption.
In this paper, we propose a novel modular DFR model that is suitable for fully digital implementations.
The proposed model reduces the number of hyperparameters and allows flexibility in the selection of the nonlinear function,
which improves the accuracy while reducing the power consumption.
We further present two DFR realizations with different nonlinear functions,
achieving 10x power reduction and 5.3x throughput improvement while maintaining equal or better accuracy.
\end{abstract}




\maketitle


\section{Introduction}\label{sec:intro}

Reservoir computing (RC) is a
machine learning method that uses a reservoir
to nonlinearly transform inputs into high-dimensional vectors.
In RC, the reservoir weights are not
altered and the weights of the output layer that follows the reservoir are the target of learning~\cite{lukovsevivcius2009reservoir},
which allows efficient training.
RC is considered suitable for time series processing because of its recurrent structure, which reflects past inputs.
Because the weights of a reservoir are fixed, it can be implemented in hardware utilizing various physical phenomena.

A delayed feedback reservoir (DFR)~\cite{appeltant2011information} is a specific type of RC system.
It is particularly suitable for hardware implementations
because it can be compactly constructed with a single nonlinear element and
a feedback loop~\cite{tanaka2019recent}.
Until now, hardware implementations of DFRs have been of two types: analog and digital~\cite{appeltant2011information}.
In an analog implementation, only the nonlinear element of the reservoir or the nonlinear element and the feedback loop
are implemented in an analog manner.
However, the inputs and outputs are generally processed digitally, requiring a digital-to-analog converter (DAC) and an analog-to-digital converter (ADC).
In addition, the time required for signal propagation through the feedback loop reduces the throughput.
In comparison, digital implementations mimic
the nonlinear behaviors of analog implementations in a fully digital manner.
However, typical implementations involve division and power calculations that require significant hardware resources.

This paper proposes a novel interpretation of the DFR equation and a corresponding DFR model
that enables an efficient fully digital implementation.
The proposed interpretation relaxes the selection of the nonlinear
function without increasing the number of hyperparameters,
thereby simplifying parameter search and improving its efficiency.
The proposed model is different from existing analog implementations, which require
a DAC and an ADC, and which involve a delay associated with the feedback loop.

Based on this model, we propose
identity and tanh~DFRs, which use
identity and hyperbolic tangent functions, respectively, as the pluggable nonlinear function.
The identity~DFR requires
neither complex modules such as division
nor exponentiation calculations,
different from existing digital implementations.
The tanh~DFR provides flexibility to adjust the nonlinearity of the reservoir.
In this study, software evaluation
validated the time series prediction accuracy compared with those of nine existing machine learning methods.
The tanh~DFR was observed to outperform these state-of-the-art methods, and the identity~DFR
demonstrated comparable accuracy to them on most datasets.
In addition, the identity~DFR was evaluated in a hardware implementation.
The hardware evaluation using a field-programmable gate array (FPGA)
showed 10x and 5.3x improvements simultaneously in the throughput and the power consumption, respectively,
compared to those of a hybrid FPGA--aplication-specific integrated circuit (ASIC) DFR~\cite{shears2021hybrid}.

The important contributions of this study can be summarized as follows:
\begin{enumerate}
 \item We developed a new interpretation of the DFR equation with a corresponding model called modular~DFR that facilitates efficient design of a fully digital DFR\@.
 \item As practical examples, we developed identity and tanh~DFRs, which use identity and hyperbolic tangent functions, respectively, as the pluggable nonlinear function.
 \item The effectiveness of the proposed DFRs was verified by comparing them with existing machine learning methods and DFR implementations using an FPGA\@.
\end{enumerate}

The remainder of this paper is organized as follows.
First, Section~\ref{sec:reservoir} reviews the concept and operation of RC and DFR\@.
Section~\ref{sec:propose} proposes a new DFR model that enhances
flexible design of a fully digital DFR and presents two practical examples of the model. 
Section~\ref{sec:evaluation} presents the evaluation of the proposed model from both software and hardware perspectives,
and, finally, Section~\ref{sec:conclusion} concludes the paper.


\section{RC}\label{sec:reservoir}

\subsection{Concept of RC}\label{subsec:RC}

RC consists of an input layer, a reservoir, and an output layer~\cite{lukovsevivcius2009reservoir}.
The input layer converts an input signal into a form that the reservoir can process.
The reservoir subsequently transforms the signal from the input layer into a high-dimensional nonlinear vector
called the \emph{reservoir state.}
Finally, the output layer performs pattern recognition based on a simple learning algorithm.
Each layer is characterized by weights connecting the nodes in that layer.
The weights of the input and reservoir layers are fixed,
and the learning process only determines the weight parameters of the output layer.

RC is considered particularly suitable for processing time-series data because of its recurrent
structure, which can reflect past inputs.
In addition, the reservoir layer does not require learning and
therefore can be implemented in fixed hardware utilizing
various physical phenomena~\cite{tanaka2019recent}.

\subsection{DFR}\label{subsec:DFR}

\begin{figure}[tb]
 \centering
 \includegraphics[width=0.8\linewidth]{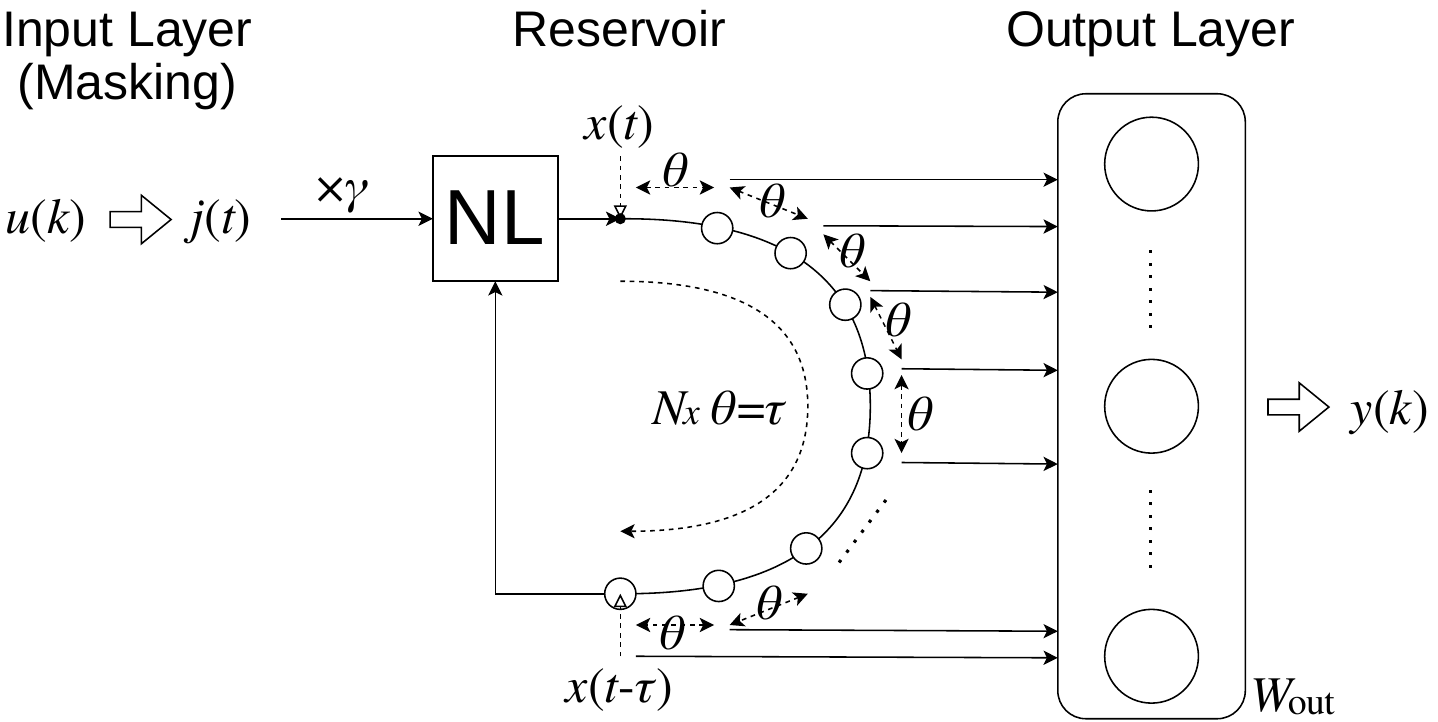}
 \caption{Conceptual diagram of DFR~\cite{appeltant2011information}.
 The reservoir consists of a nonlinear element (NL) and a feedback loop with a total delay $\tau$.
 The feedback loop comprises $N_x$ virtual nodes with a time interval $\theta$.}\label{fig:DFR}
\end{figure}

Figure~\ref{fig:DFR} illustrates a conceptual diagram of a DFR\@.
First, the input signal, $u(k)$, is time-multiplexed to $j(t)$ by masking and is input to the nonlinear element (NL)
with the feedback signal.
The nonlinear element is represented by the following delay differential equation:
\begin{equation}
 \frac{\mathrm d}{{\mathrm d}t}x(t)=F(x(t-\tau), \gamma j(t)).
 \label{eq:delay}
\end{equation}
Consider that the virtual nodes are connected at a time interval $\theta$ on
the feedback loop with a total delay $\tau$.
The signal values at the virtual nodes are the features, and they are collected as a vector
of $N_x$ elements as the reservoir state.
Here, the following relationship holds:
\begin{equation}
 N_x\theta =\tau.
 \label{eq:nx}
\end{equation}
The reservoir state is defined as follows:
\begin{equation}
 \boldsymbol{x}(k)\equiv[x(k\tau -\theta ), x(k\tau -2\theta ), \ldots , x(k\tau -\tau)].
 \label{eq:x}
\end{equation}
The output of the DFR is obtained by
linearly transforming the reservoir state at the output layer.
The detailed operation is explained subsequently.

\subsection{Analog implementation of DFR}\label{subsec:analogDFR}

The reservoir in the DFR has a relatively simple structure consisting of a nonlinear element and a feedback loop.
Conversely, well-known RC methods, such as the echo state network, employ a recurrent neural network as the reservoir.
A DFR is easier to implement in hardware than other RC methods~\cite{tanaka2019recent}.
Most existing hardware implementations of DFRs use analog circuits in the reservoir~\cite{appeltant2011information,soriano2014delay}.
\begin{figure}[tb]
 \centering
 \includegraphics[width=0.7\linewidth]{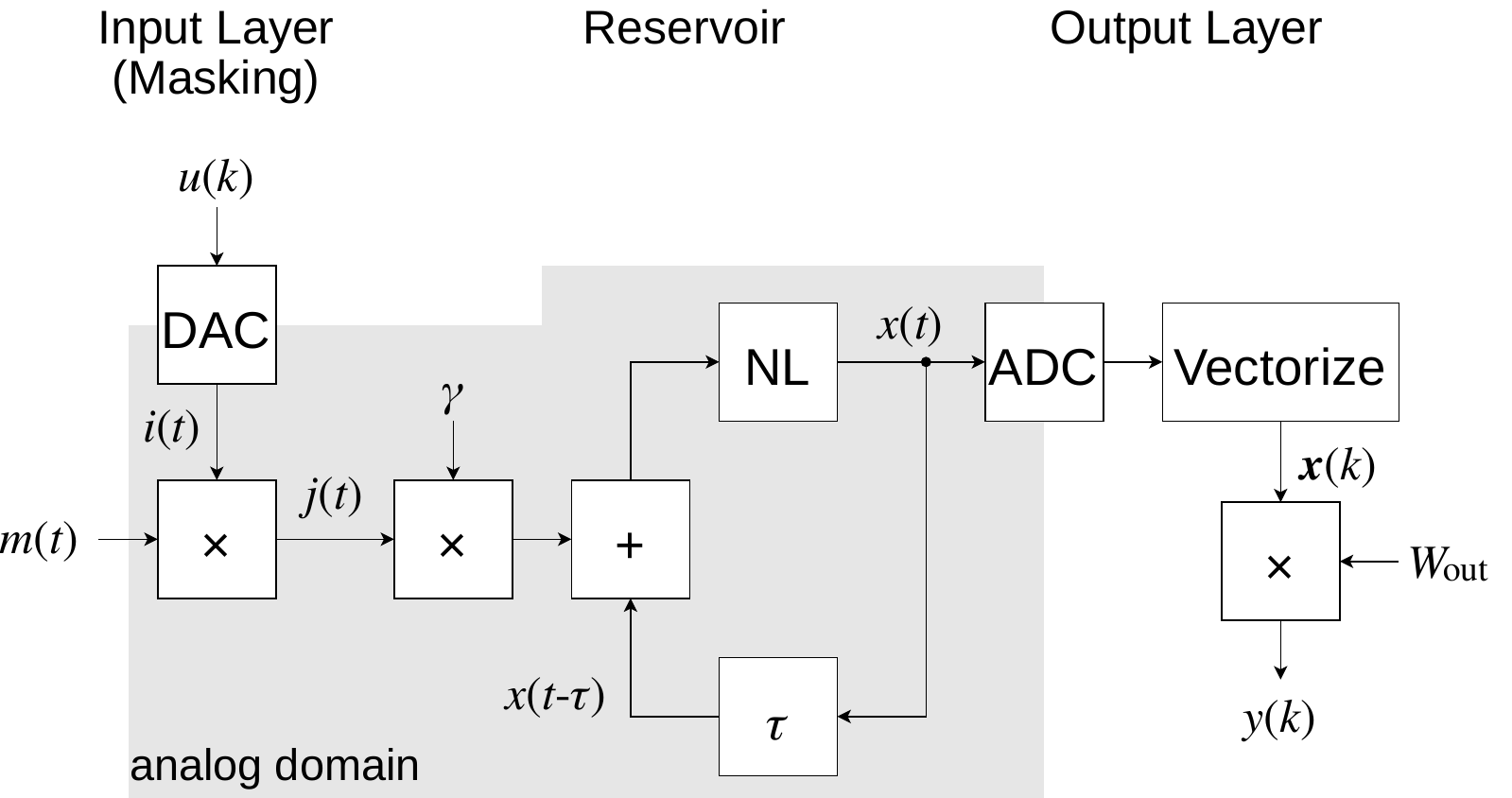}
 \caption{Block diagram of DFR for analog implementation of a nonlinear element and a feedback loop~\cite{appeltant2011information}.
 The DAC and ADC are placed before masking and after the reservoir, respectively.}\label{fig:analogDFR}
\end{figure}

A block diagram of an analog implementation of a DFR is shown in Fig.~\ref{fig:analogDFR}.
First, the digital time series input, $u(k)$, sampled with a period $\tau$,
is converted to an analog signal $i(t)$.
\begin{figure}[tb]
 \centering
 \includegraphics[width=0.8\linewidth]{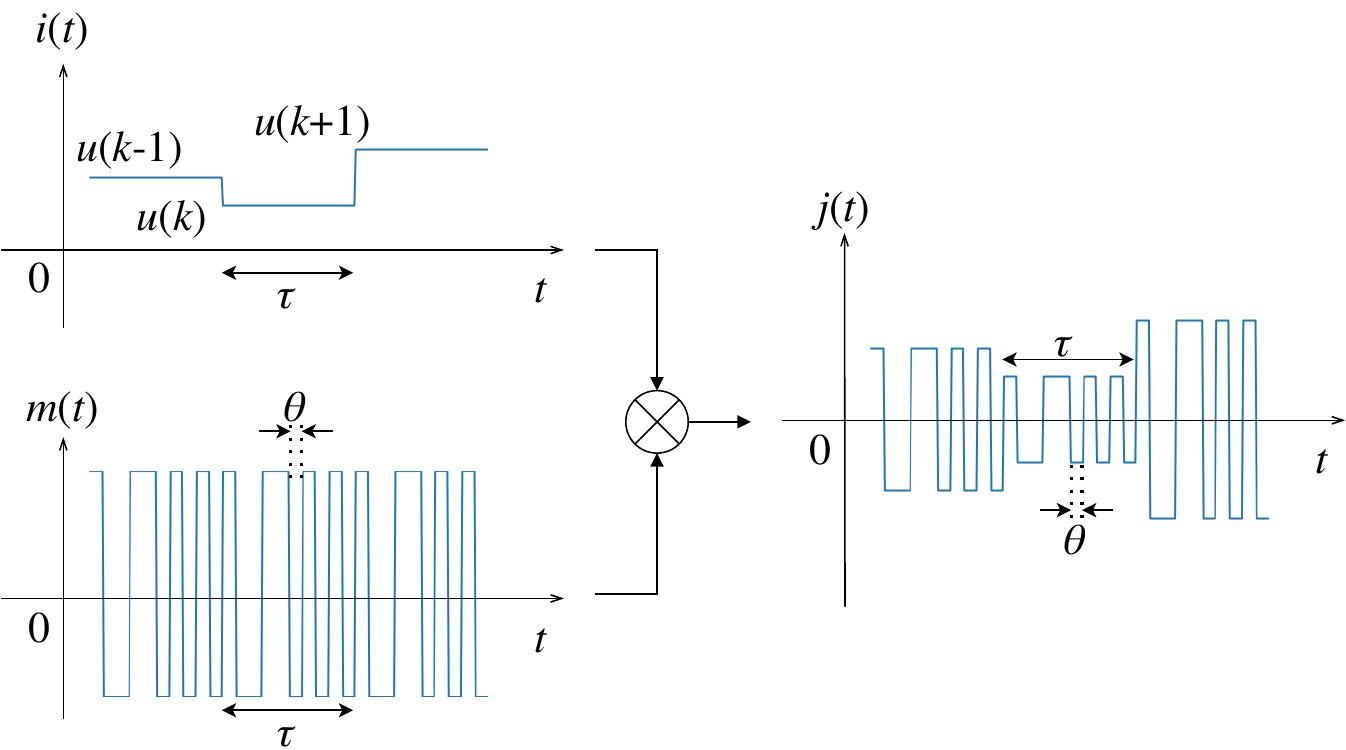}
 \caption{
 Masking process. $i$ is obtained by digital-to-analog conversion of input signal $u$.
 Signal $i$ is constant at all $\tau$.
 Mask signal $m$ takes different values at all time intervals $\theta$, and its period is $\tau$.
 Input signal to reservoir is expressed as $j(t) = i(t) \cdot m(t)$.
}\label{fig:mask}
\end{figure}
$i(t)$ is subsequently multiplied by the mask signal with a faster sample rate $\theta$ in the masking process (Fig.~\ref{fig:mask}).
Here, $\theta$ is the delay between the nodes in the reservoir.
The length of the mask pattern is equal to the number of virtual nodes in the DFR, as expressed in Eq.~(\ref{eq:nx}).
Each mask value is applied at each time interval $\theta$.
Fig.~\ref{fig:mask} shows the most common mask signals, $+1$ and $-1$; however,
different masks can be used~\cite{appeltant2011information,appeltant2014constructing}.

$j(t)$ is subsequently scaled by $\gamma$, added to the reservoir feedback, and enters
the nonlinear element (NL), which produces the output, $x(t)$.
Mackey--Glass (MG) model~\cite{mackey1977oscillation} is commonly used as the NL~\cite{appeltant2011information,soriano2014delay,alomar2015digital}.
Its operation is expressed as follows:
\begin{eqnarray}
 \frac{\mathrm d}{{\mathrm d}t}x(t)=-x(t)+\eta f(x(t-\tau),\gamma j(t)),\\
 \label{eq:MG}
 f(x(t-\tau),\gamma j(t))=\frac{[x(t-\tau)+\gamma j(t)]}{1+{[x(t-\tau)+\gamma j(t)]}^p},
 \label{eq:f}
\end{eqnarray}
where $p$ is an adjustable parameter.
$x(t)$ is sampled and converted to digital at each time interval $\theta$.
For each input data, sampling is performed $N_x$ times, and the reservoir state, or the feature vector, is
obtained, as expressed in Eq.~(\ref{eq:x}).
Simultaneously, $x(t)$ undergoes the feedback loop with a delay time $\tau$ and is added to $\gamma j(t)$.
Note that Eq.~(\ref{eq:nx}) holds.

The output is obtained by linearly transforming the reservoir state at the output layer.
In~\cite{ikeda2022hardware}, the reservoir state, $\boldsymbol{\tilde{x}}(k)$, is defined as follows:
\begin{equation}
 \boldsymbol{\tilde{x}}(k)\equiv[\boldsymbol{x}(k),1].
\end{equation}
The output is subsequently obtained by applying a linear transformation as follows:
\begin{equation}
 y(k)=W_{\mathrm{out}}\boldsymbol{\tilde{x}}(k).
 \label{eq:linear}
\end{equation}

Various circuit designs with different ratios of analog to digital circuits exist.
Examples include a primarily digital approach in which only the nonlinear
element is implemented in analog,
whereas the feedback loop is in the digital domain~\cite{appeltant2011information,shears2021hybrid}.
In this approach, the DAC and the ADC in a DFR are placed before and after the nonlinear element, respectively.
One common aspect of these implementations is that both DAC and ADC are essential.
Although digital circuits may consume less power,
power reduction of the converter circuits is significantly challenging.
Additionally, obtaining an element of the reservoir state requires time $\theta$,
and obtaining the reservoir state for each input takes time $\tau$.
The conversion times of the ADC and the DAC further reduce the throughput.

\subsection{Digital implementation of DFR}\label{subsec:digitalDFR}

Fully digital implementations of a DFR also exist~\cite{appeltant2011information,alomar2015digital}.
In digital DFRs, the signal, $\boldsymbol{j}(k)$, after the masking process is expressed as~\cite{appeltant2011information}
\begin{equation}
 \boldsymbol{j}(k)=\boldsymbol{m}u(k),
\end{equation}
where $\boldsymbol{m}$ is a mask vector of length $N_x$
and whose elements are determined randomly and digitally.
After masking, the nonlinear element and feedback loop of the DFR determine the reservoir state, $\boldsymbol{x}(k)$.

Most implementations use the MG model as the nonlinear element.
Assuming that $f$ is constant for a short time $\theta$, which is the time interval between the virtual nodes,
the differential equation in Eq.~(\ref{eq:MG}) is solved for $x(t)$ as follows:
\begin{equation}
 x(t)=x_0\e^{-t}+\eta (1-\e^{-t})f(x(t-\tau)+\gamma j(t)),
\end{equation}
where $x_0$ is the initial value of $x(t)$ at each $\theta$~\cite{appeltant2011information}.
The value of the next virtual node can be expressed as $x(\theta)$.
Assuming that the components of $\boldsymbol{x}(k)$ and $\boldsymbol{j}(k)$ are
\begin{eqnarray}
 \boldsymbol{x}(k)&\equiv&[x(k)_1,x(k)_2,\ldots,x(k)_{N_x}],\\
 \boldsymbol{j}(k)&\equiv&[j(k)_1,j(k)_2,\ldots,j(k)_{N_x}],
\end{eqnarray}
$\boldsymbol{x}(k)$ is obtained recurrently as follows:
\begin{align}
 \label{eq:digitalDFR1}
  {x(k)}_1&=\ x(k-1)_{N_x}\e^{-\theta} +\eta (1\!-\!\e^{-\theta})f(x(k-1)_1+\gamma j(k)_1),\\
 \label{eq:digitalDFR2}
  {x(k)}_n&=\ x(k)_{n-1}\e^{-\theta}   +\eta (1\!-\!\e^{-\theta})f(x(k-1)_n+\gamma j(k)_n).\ (n\geq2)
\end{align}
Here, the initial value of the reservoir state is set to 0.
The construction and operation of the output layer are the same as in an analog implementation.

Although the existing digital implementations avoid the use of a DAC and an ADC\@, 
the choice of the nonlinear function, $f$, is limited to those mimicking analog implementations,
such as the MG model, which requires significant hardware resources.


\section{New Digital DFR Model Enhancing Design Flexibility}\label{sec:propose}

In this section, we propose a new DFR model suitable for digital implementations.
The primary objective in its development was to improve the flexibility of choosing the nonlinear block in a digital DFR\@.
The freedom to choose the nonlinear block simplifies the design and enhances the accuracy compared with existing digital DFRs.
Although adding this extra design feature may seem to increase the design complexity,
in this section, we subsequently show analytically that the hyperparameter, $\gamma$, for input scaling can be eliminated,
which eventually simplifies the overall design. 
Finally, we present two examples of the new DFR model.

\subsection{Modular DFR}\label{subsec:newmodel}

\begin{figure}[tb]
 \centering
 \includegraphics[width=0.75\linewidth]{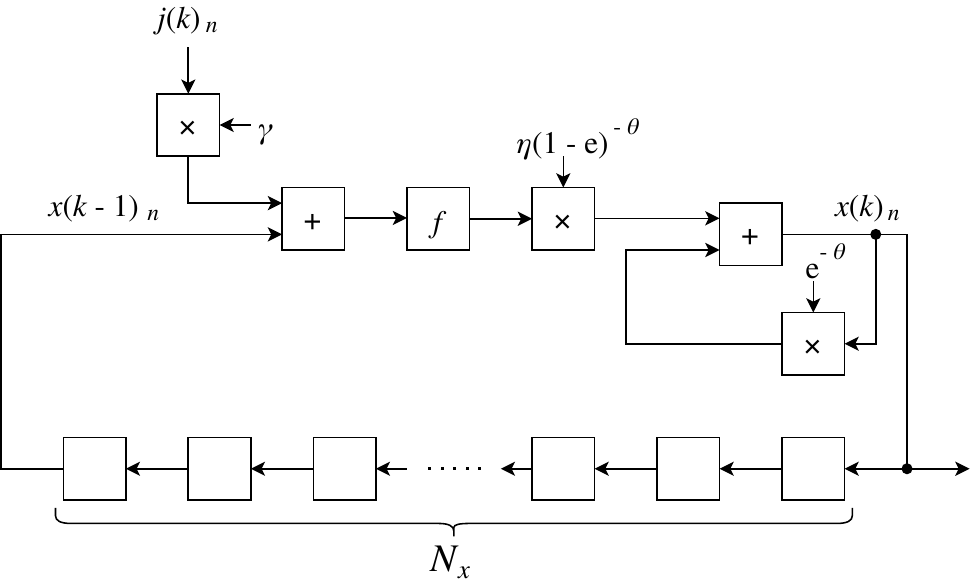}
 \caption{
 Block diagram of reservoir processing based on the reformulation of the traditional digital DFR model.
 Block labeled ``$f$'' operates according to one-input, one-output function $f$ in Eq.~(\ref{eq:MG}).
 Empty blocks are virtual nodes of the reservoir, which functions as shift registers.
}\label{fig:digitalDFR}
\end{figure}

Based on Eq.~(\ref{eq:digitalDFR2}),
${x(k)}_n$ depends on ${x(k)}_{n-1}$, ${x(k-1)}_n$, and ${j(k)}_n$.
The recursive relationship between the inputs, ${x(k)}_{n-1}$
and ${x(k-1)}_n$, with the output ${x(k)}_n$
poses a challenge for designing a digital nonlinear block of a DFR reservoir.
To address this issue, we propose
reformulation of the traditional model to use a one-input, one-output function $f$.
The functionality of a reservoir in the digital domain can be represented, as shown in Fig.~\ref{fig:digitalDFR}.
This interpretation simplifies the design process by decomposing the operations of the nonlinear
element of the reservoir into multiple blocks.
This makes the design of the one-input, one-output function, $f$, independent of other parameters.

The original DFR reservoir requires adjustment of three parameters: $\gamma$, $\eta$, and $\theta$.
To mitigate the design complexity due to the flexibility in choosing the nonlinear function,
we conducted an analytical study of the relationships between these parameters.
We supposed that $\gamma$ is scaled to $\alpha \gamma$ and $x'$ is obtained accordingly.
Furthermore, we assumed that the scaling relationship holds for the reservoir state at the previous time, $(k-1)$, expressed as follows:
\begin{equation}
 \boldsymbol{x}'(k-1) = \alpha \boldsymbol{x}(k-1).
 \label{eq:condition}
\end{equation}
We consider the changes in the variable, $x'$, and scaling of $f$ to define function $f'$ as follows:
\begin{equation}
 f'(x)=\alpha f(\frac{x}{\alpha}).
 \label{eq:fprime}
\end{equation}
Therefore, the first virtual node at time $k$ satisfies scaling by $\alpha$ as follows:
\begin{align}
 {x'(k)}_1 = &\ {x'(k-1)}_{N_x}\e^{-\theta}+\eta (1-\e^{-\theta})f'({{x'(k-1)}_1}+\alpha \gamma {j(k)}_1)&\nonumber\\
           = &\ \alpha {x(k-1)}_{N_x}\e^{-\theta}+\alpha \eta (1-\e^{-\theta})f({x(k-1)}_1+\gamma {j(k)}_1)&\nonumber\\
           =&\ \alpha {x(k)}_1.
\end{align}
A similar calculation yielded that the scaling relationship, ${x'(k)}_n=\alpha {x(k)}_n\ (n\geq2)$, holds for all other nodes.
Because the relation in Eq.~(\ref{eq:condition}) also holds for $k=1$ with zero as the initial value of $\boldsymbol{x}'$, by induction,
this relation holds for all $k$ as follows:
\begin{equation}
 \boldsymbol{x}'(k) = \alpha \boldsymbol{x}(k).
 \label{eq:xprime}
\end{equation}
Subsequently, the reservoir state produced by any input scaling with $\gamma$ can be achieved equivalently
by substituting $f$ with $f'$ in Eq.~(\ref{eq:fprime}).

\begin{figure}[tb]
 \centering
 \includegraphics[width=0.8\linewidth]{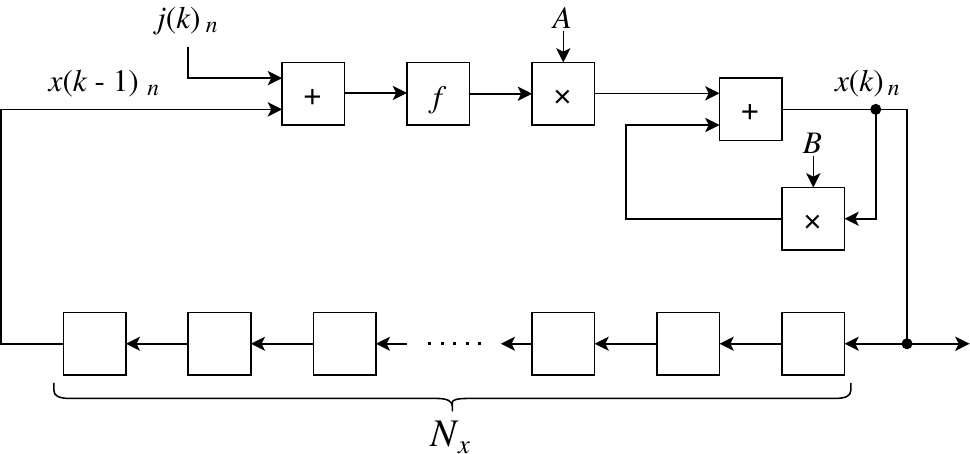}
 \caption{Block diagram of the proposed modular DFR\@.
 Only two parameters, $A$ and $B$, have to be optimized.}\label{fig:newDFR}
\end{figure}

The above discussion also applies to the output layer.
Ridge regression, a widely used method for training the output layer of a DFR, is performed as follows.
First, let $d(k)$ be the target output for an input $u(k)$.
We subsequently define the following:
\begin{eqnarray}
 \tilde{X}&=&[\tilde{x}(1),\tilde{x}(2),\ldots,\tilde{x}(T)],\\
 D&=&[d(1),d(2),\ldots,d(T)],
\end{eqnarray}
where $T$ is the number of training data.
In ridge regression, $W_{\mathrm{out}}$ is derived as follows:
\begin{equation}
 W_{\mathrm{out}}=D\tilde{X}^{\mathrm{T}}(\tilde{X}\tilde{X}^{\mathrm{T}}+\beta I)^{-1},
 \label{eq:ridge}
\end{equation}
where $I$ is an identity matrix of size $N_x \times N_x$ and $\beta$ is a regularization parameter.
Once more, considering the scaling of the input, $\tilde{X}$, and parameter $\beta$:
\begin{eqnarray}
 \tilde{X}'=\alpha \tilde{X},\\
 \beta '=\alpha^2 \beta,
\label{eq:square}
\end{eqnarray}
the corresponding $W_{\mathrm{out}}'$ to the input scaling is obtained as
\begin{eqnarray}
 W_{\mathrm{out}}'&=&D\tilde{X}'^{\mathrm{T}}(\tilde{X}'\tilde{X}'^{\mathrm{T}}+\beta' I)^{-1}\nonumber\\
 &=&D\alpha \tilde{X}^{\mathrm{T}}(\alpha^2 \tilde{X}\tilde{X}^{\mathrm{T}}+\alpha^2 \beta I)^{-1}\nonumber\\
 &=&\frac{1}{\alpha} W_{\mathrm{out}}.
  \label{eq:woutp}
\end{eqnarray}
Because the output, $y(k)$, is derived using Eq.~(\ref{eq:linear}), the same result as the input scaling can be achieved
using the matrix in Eq.~(\ref{eq:woutp}) with $\beta$ multiplied by $\alpha^2$.
In summary, using the new model, called modular DFR in Fig.~\ref{fig:newDFR}, the same solution space
with the original DFR is obtained.
The proposed model reduces the parameter space to be explored to two dimensions while enabling the selection of function $f$,
making the optimization of a digital DFR easier and more efficient.
$A$ and $B$ in Fig.~\ref{fig:newDFR} correspond to $\eta(1-\e^{-\theta})$ and $\e^{-\theta}$ in Fig.~\ref{fig:digitalDFR}.

The most crucial aspect of a DFR design is the selection of the nonlinear element and associated parameters.
In an analog implementation, selecting an element that satisfies the delay differential equation is sufficient.
The proposed model simplifies the digital DFR design process, making it as straightforward as its analog counterpart.
Instead of reproducing the recursive differential equation, we can more freely design
the one-input, one-output function, $f$, with fewer parameters to learn.
In addition, the processing in the proposed modular DFR, except that for $f$, is simple,
comprising only constant multiplications, additions, and shifts.
This makes it easier to implement low-power and high-throughput digital DFR using existing design tools.

\subsection{Examples of proposed model}\label{subsec:propose}

As examples of the proposed modular DFR, we present two DFRs that use identity or hyperbolic tangent functions as $f$.

\begin{figure}[tb]
 \centering
 \includegraphics[width=0.8\linewidth]{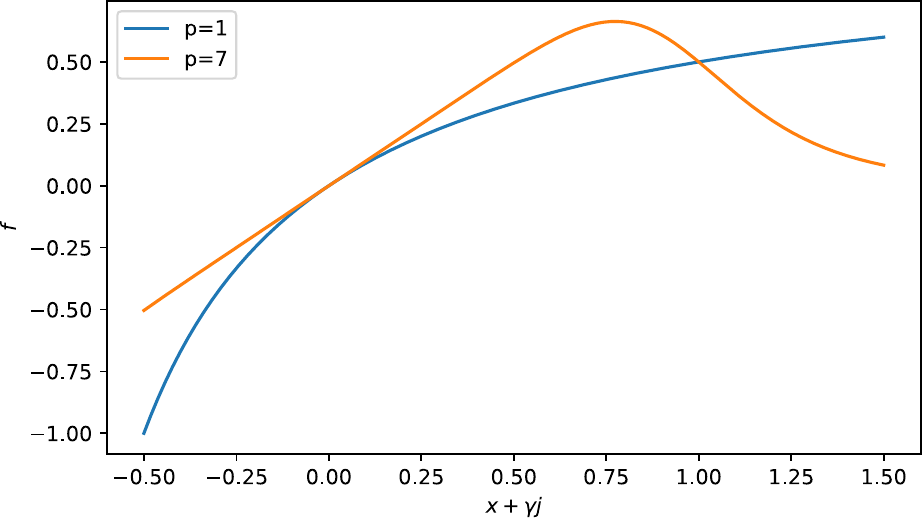}
 \caption{
 Behavior of $f$ in Mackey--Glass model.
 Parameter $p$ for adjusting nonlinearity are 1 and 7, as used in~\cite{appeltant2011information}.
 }\label{fig:MG_nl}
\end{figure}

\begin{figure}[tb]
 \centering
 \includegraphics[width=0.8\linewidth]{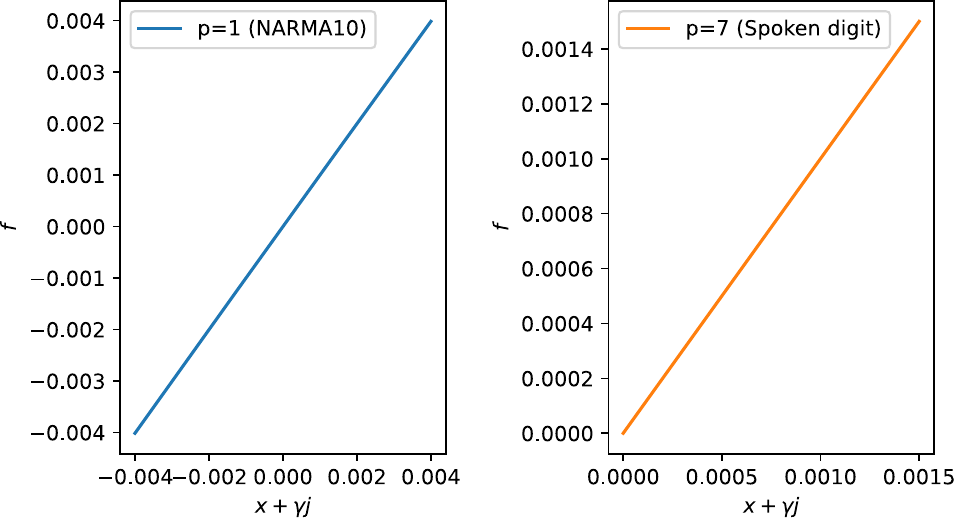}
 \caption{
 Output of function $f$ with input range used to solve tasks in~\cite{appeltant2011information}.
 $p=1$ and $p=7$ correspond to NARMA10 and spoken digit tasks, respectively.
 Only weak nonlinearity was used in both examples.
 }\label{fig:MG_range}
\end{figure}

Fig.~\ref{fig:MG_nl} shows the shapes of the MG models as $f$, with $p=1$ and $p=7$ as used in~\cite{appeltant2011information}.
The function, $f$, exhibits nonlinearity, the degree of which can be adjusted with $p$.
\cite{appeltant2011information} evaluated the performance of a DFR with the MG model using NARMA10 and spoken digit,
which are time-series prediction and spoken word classification tasks, respectively.
The ranges of $x+\gamma j$ to solve the tasks are depicted in Fig.~\ref{fig:MG_range} using the same parameters as in~\cite{appeltant2011information}.
Notably, although $f$ in the MG model exhibits nonlinearity,
only a narrow region, which is considered to be mostly linear, is used in most tasks, including those mentioned above.

\begin{figure}[tb]
 \centering
 \includegraphics[width=0.8\linewidth]{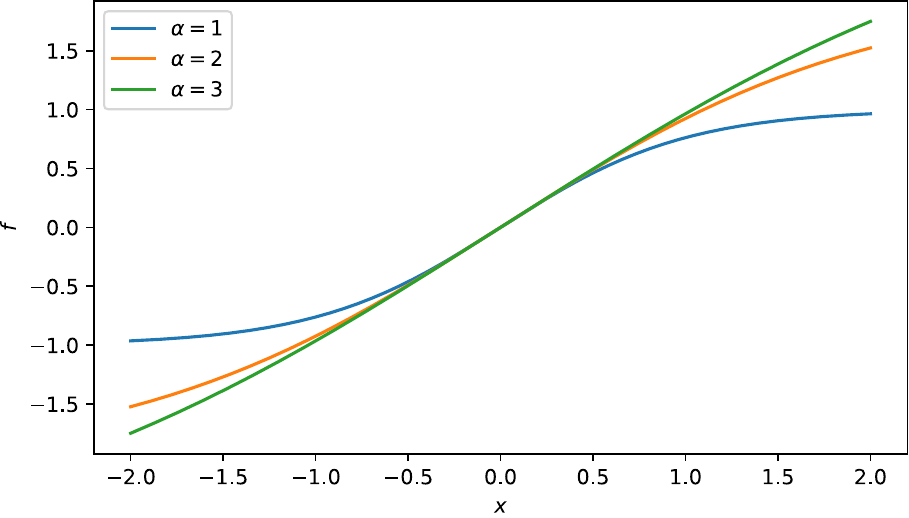}
 \caption{
 Input--output characteristics of the hyperbolic tangent function as $f$ (Eq.~(\ref{eq:tanh})) for different $\alpha$.
 }\label{fig:tanh}
\end{figure}

Based on this rationale, we propose two examples of the nonlinear function, $f$, of the developed modular DFR\@.
Because both these examples are characterized by a linear region near the origin,
we propose an identity function as $f$ as the first and simplest example.
In this case, since $f(x)=x$ holds, the scaling of $f$ expressed in Eq.~(\ref{eq:fprime}) is unnecessary.
The second example is a hyperbolic tangent function as $f$ to better control the nonlinearity, which is expressed as:
\begin{equation}
 f(x)=\alpha\tanh\frac{x}{\alpha}=\alpha\frac{\e^{\frac{x}{\alpha}}-\e^{-\frac{x}{\alpha}}}{\e^{\frac{x}{\alpha}}+\e^{-\frac{x}{\alpha}}}.
 \label{eq:tanh}
\end{equation}
This formula considers the parameter scaling as expressed in Eq.~(\ref{eq:fprime}).
The shape of $f$ when it is a hyperbolic tangent function for different $\alpha$ is plotted in Fig.~\ref{fig:tanh}.
Increasing and dicreasing the value of $\alpha$ widen and narrow the linear section, respectively.
Thus, by changing $\alpha$, the nonlinearity of $f$ can be adjusted.


\section{Evaluation}\label{sec:evaluation}

In this section, we present the performance evaluation of the proposed DFR models.
First, we compare their accuracy with those of existing machine learning techniques.
Subsequently, existing hardware implementations of
DFRs are compared in terms of the accuracy, circuit size, power consumption, and throughput.

\subsection{Comparison with existing machine learning techniques}
To compare the proposed methods with existing machine learning techniques, we employed three time-series prediction tasks.
Two of these involved typical chaotic systems: the Lorenz and R\"ossler systems.
The third task was a real magnetic storm loop current index dataset,
specifically the disturbance storm time (DST) dataset~\cite{banerjee2012existence}.
These tasks were the same as those used in the state-of-the-art chaotic time series prediction method, MOICBLS~\cite{yi2022intergroup}.

The Lorenz system is defined as:
\begin{eqnarray}
 \frac{{\mathrm d}x}{{\mathrm d}t}&=&-a(x-y),\nonumber \\
 \frac{{\mathrm d}y}{{\mathrm d}t}&=&-xz+cx-y,\\
 \frac{{\mathrm d}z}{{\mathrm d}t}&=&xy-bz,\nonumber
\end{eqnarray}
where $a=10$, $b=8/3$, $c=28$, $x(0)=1$, $y(0)=0$, and $z(0)=1$ in the following experiments.

The R\"ossler system is defined as:
\begin{eqnarray}
 \frac{{\mathrm d}x}{{\mathrm d}t}&=&-y-z,\nonumber \\
 \frac{{\mathrm d}y}{{\mathrm d}t}&=&x+ay,\\
 \frac{{\mathrm d}z}{{\mathrm d}t}&=&b+z(x-c),\nonumber
\end{eqnarray}
where $a=0.15$, $b=0.2$, $c=10.0$, $x(0)=1$, $y(0)=0$, and $z(0)=1$.

For both systems, the fourth-order Runge--Kutta method was used to generate 12,000 samples with a sampling interval of 0.01~s.
Of these, the first 9,000 samples were used for training, and the remaining 3,000 were used for testing.

The DST index dataset, which has a sampling period of 1 h~\cite{DST}, was used in this study.
Specifically, 4,000 samples from the tail of both the 2013 and 2014 data were used.
Among these, the first 3,000 samples were allocated for training,
whereas the remaining 1,000 were used for testing.
Note that all the settings used in this study were consistent with those in~\cite{yi2022intergroup}.

The machine learning methods compared were as follows:
\begin{enumerate}
\item random forest (RF)~\cite{xue2019semg}
\item extreme learning machine (ELM)~\cite{yan2017fast}
\item support vetor machine (SVM)~\cite{chen2015weighted}
\item elman artificial neural network (Elman)~\cite{chandra2015competition}
\item denoising autoencoder (DAE)~\cite{laumanns2002combining}
\item gradient boosting (GB)~\cite{guo2019parallel}
\item echo state network (ESN)~\cite{han2017laplacian}
\item broad learning system (BLS)~\cite{chen2017broad}
\item intergroup cascade BLS with multiobjective optimized parameters (MOICBLS)~\cite{yi2022intergroup}
\end{enumerate}
The prediction accuracies of these methods, as reported in~\cite{yi2022intergroup} were used.

In addition, three DFRs were evaluated: a general DFR that uses the MG model as the nonlinear element (MG~DFR),
the proposed modular DFR that uses an identity function as $f$ (identity~DFR)
and the proposed modular DFR that uses a hyperbolic tangent function as $f$ (tanh DFR).
Python was used to obtain the prediction accuracy of the DFRs.
The mask signal had two values, $+1$ and $-1$, and the number of virtual nodes $N_x$ was 100 for all reservoirs.
In the MG~DFR, fixed parameters $p=1$ and $\theta=0.2$ were used for all datasets.
For the identity and tanh~DFRs, we used $B=0.82$.
Here, $\theta=0.2$ is the value at which the DFR is considered to work effectively because coupling between the virtual nodes is optimal~\cite{appeltant2011information}.
Because $B=\e^{-\theta}$, fixing $\theta=0.2$ was equivalent to fixing $B=0.82$.
Other parameters were optimized by grid search, and the values used are summarized in Table~\ref{tab:param1}.
Note that the weights of the linear transform in the output layer were learned by ridge regression using the regularization parameter, $\beta$.
For each dataset, normalization was implemented with the maximum value set to 1.
For the identity and tanh~DFRs, $\beta$ needed to be multiplied by $(1/\mathrm{max}_k{u(k)})^2$.
\begin{table}
 \centering
 \caption{Parameters of the DFRs used for comparison with existing machine learning techniques.}\label{tab:param1}
 \begin{tabular}{c|ccc|cc|ccc} \hline
  dataset & \multicolumn{3}{c|}{MG~DFR} & \multicolumn{2}{c|}{identity~DFR} & \multicolumn{3}{c}{tanh~DFR}\\
  & $\gamma$ & $\eta$ & $\beta$ & $A$ & $\beta$ & $A$ & $\alpha$ & $\beta$ \\ \hline\hline
  Lorenz~$x$ & 0.02 & 0.001 & 1E-17 & 0.09 & 1E-5 & 0.1 & 10 & 1E-5 \\ \hline
  Lorenz~$y$ & 0.04 & 0.5 & 1E-11 & 0.05 & 1E-7 & 0.15 & 4 & 1E-4 \\ \hline
  Lorenz~$z$ & 0.02 & 0.04 & 1E-14 & 0.1 & 1E-6 & 0.07 & 15 & 1E-6 \\ \hline
  R\"{o}ssler~$x$ & 0.003 & 0.01 & 1E-16 & 0.04 & 1E-4 & 0.04 & 4000 & 1E-4 \\ \hline
  R\"{o}ssler~$y$ & 0.001 & 0.001 & 1E-20 & 0.01 & 1E-6 & 0.02 & 400 & 1E-5 \\ \hline
  R\"{o}ssler~$z$ & 0.06 & 0.5 & 1E-12 & 0.04 & 1E-8 & 0.1 & 10 & 1E-6 \\ \hline
  DST~2013 & 3E-4 & 0.3 & 1E-18 & 0.003 & 1E-6 & 0.04 & 2 & 1E-5 \\ \hline
  DST~2014 & 4E-4 & 0.35 & 1E-18 & 0.003 & 1E-6 & 0.08 & 4 & 1E-6 \\ \hline
 \end{tabular}
\end{table}

The results of the accuracy evaluation on the Lorentz time series are summarized in Table~\ref{tab:Lorenz}.
To evaluate the accuracy of the proposed methods and compare them with existing ones,
we used the root mean square error (RMSE) and the mean absolute error (MAE), as in~\cite{yi2022intergroup}.
The RMSE and MAE were defined based on the predicted output ($y_i$), target output ($d_i$),
and number of samples in the test data ($n$) as follows:
\begin{eqnarray}
 \mathrm{RMSE}&=&\sqrt{\frac{1}{n}\sum_{i=1}^{n}{(d_i-y_i)^2}},\\
 \mathrm{MAE}&=&\frac{1}{n}\sum_{i=1}^{n}{|d_i-y_i|}.
\end{eqnarray}
\begin{table}
 \centering
 \caption{Prediction results of Lorenz time series.
 Values except for DFRs are obtained from~\cite{yi2022intergroup}.
 The best and second-best values for each dataset are shown in bold and underlined, respectively.}
 \label{tab:Lorenz}
 \begin{tabular}{c|ccc|ccc} \hline
  & \multicolumn{3}{c|}{RMSE} & \multicolumn{3}{c}{MAE}\\
  methods & $x$ & $y$ & $z$ & $x$ & $y$ & $z$ \\ \hline\hline
  RF & 3.504E-4 & 5.265E-4 & 7.820E-4 & 3.363E-4 & 6.028E-4 & 6.890E-4 \\ \hline
  ELM & 9.113E-4 & 2.226E-3 & 1.968E-3 & 8.746E-4 & 2.515E-3 & 1.768E-3 \\ \hline
  SVM & 4.525E-3 & 5.774E-3 & 5.553E-2 & 4.342E-3 & 6.140E-3 & 5.471E-2 \\ \hline
  Elman & 2.867E-4 & 5.481E-4 & 6.746E-3 & 2.752E-4 & 5.679E-4 & 6.402E-4 \\ \hline
  DAE & 1.017E-3 & 2.384E-3 & 2.013E-3 & 9.765E-4 & 2.686E-3 & 1.994E-3 \\ \hline
  GB & 3.237E-4 & 5.607E-4 & 7.117E-4 & 3.106E-4 & 6.916E-4 & 6.855E-4 \\ \hline
  ESN & 3.122E-4 & 5.567E-4 & 6.716E-4 & 2.996E-4 & 6.023E-4 & 6.675E-4 \\ \hline
  BLS & 2.884E-4 & 5.277E-4 & 6.299E-4 & 2.859E-4 & 5.676E-4 & 6.245E-4 \\ \hline
  MOICBLS & 2.591E-4 & 4.089E-4 & \bf{4.766E-4} & 2.486E-4 & 4.424E-4 & \underline{4.523E-4} \\ \hline
  MG~DFR & \underline{7.996E-5} & \underline{3.219E-4} & 7.208E-4 & \underline{5.337E-5} & \underline{2.200E-4} & 5.200E-4 \\ \hline
  identity~DFR & 1.353E-4 & 3.195E-3 & 3.986E-3 & 9.059E-5 & 2.344E-3 & 2.645E-3 \\ \hline
  tanh~DFR & \bf{2.916E-5} & \bf{2.888E-4} & \underline{6.281E-4} & \bf{2.078E-5} & \bf{1.786E-4} & \bf{4.358E-4} \\ \hline
 \end{tabular}
\end{table}

The accuracy evaluation results for the R\"{o}ssler
and DST time series are presented in Tables~\ref{tab:Rossler} and \ref{tab:DST}, respectively.
\begin{table}
 \centering
 \caption{Prediction results of R\"{o}ssler time series.
 Values except for DFRs are obtained from~\cite{yi2022intergroup}.
 The best and second-best values for each dataset are shown in bold and underlined, respectively.}
 \label{tab:Rossler}
 \begin{tabular}{c|ccc|ccc} \hline
  & \multicolumn{3}{c|}{RMSE} & \multicolumn{3}{c}{MAE}\\
  methods & $x$ & $y$ & $z$ & $x$ & $y$ & $z$ \\ \hline\hline
  RF & 8.158E-4 & 7.665E-3 & 3.910E-3 & 8.363E-4 & 7.256E-3 & 3.445E-3 \\ \hline
  ELM & 1.262E-3 & 1.096E-3 & 9.844E-3 & 1.139E-3 & 1.209E-3 & 8.846E-3 \\ \hline
  SVM & 6.215E-3 & 9.075E-3 & 2.776E-1 & 5.756E-3 & 8.463E-3 & 2.735E-1 \\ \hline
  Elman & 3.867E-3 & 6.848E-4 & 3.373E-2 & 2.944E-3 & 5.679E-4 & 3.201E-3 \\ \hline
  DAE & 3.502E-3 & 7.758E-4 & 1.006E-2 & 3.765E-3 & 7.349E-4 & 9.972E-3 \\ \hline
  GB & 6.237E-4 & 8.690E-4 & 3.558E-3 & 6.106E-4 & 8.619E-4 & 3.427E-3 \\ \hline
  ESN & 5.688E-4 & 2.517E-4 & 3.357E-2 & 5.037E-4 & 2.319E-4 & 3.337E-3 \\ \hline
  BLS & 5.335E-4 & 2.327E-4 & 3.149E-2 & 4.725E-4 & 2.162E-4 & 3.111E-3 \\ \hline
  MOICBLS & 3.559E-4 & 1.210E-4 & 2.383E-3 & 3.239E-4 & 1.115E-4 & 2.226E-3 \\ \hline
  MG~DFR & 2.256E-5 & 1.443E-5 & \bf{1.391E-4} & 1.100E-5 & 1.222E-5 & \bf{9.923E-5} \\ \hline
  identity~DFR & \underline{6.069E-6} & \underline{3.851E-6} & 6.940E-4 & \underline{1.980E-6} & \underline{3.228E-6} & 4.241E-4 \\ \hline
  tanh~DFR & \bf{6.067E-6} & \bf{2.841E-6} & \underline{2.174E-4} & \bf{1.724E-6} & \bf{1.915E-6} & \underline{1.080E-4} \\ \hline
 \end{tabular}
\end{table}

\begin{table}
 \centering
 \caption{Prediction results of DST time series.
 Values except for DFRs are obtained from~\cite{yi2022intergroup}.
 The best and second-best values for each dataset are shown in bold and underlined, respectively.}
 \label{tab:DST}
 \begin{tabular}{c|cc|cc} \hline
  & \multicolumn{2}{c|}{DST~2013} & \multicolumn{2}{c}{DST~2014}\\
  methods & RMSE & MAE & RMSE & MAE \\ \hline\hline
  RF & 5.660 & 5.313 & 12.669 & 11.217 \\ \hline
  ELM & 3.866 & 3.965 & 8.529 & 7.988 \\ \hline
  SVM & 3.815 & 3.802 & 8.059 & 6.642 \\ \hline
  Elman & 4.006 & 4.028 & 8.632 & 7.798 \\ \hline
  DAE & 3.805 & 3.798 & 8.438 & 7.911 \\ \hline
  GB & 4.652 & 4.293 & 10.315 & 9.319 \\ \hline
  ESN & 3.985 & 3.905 & 8.648 & 7.912 \\ \hline
  BLS & 3.972 & 3.885 & 8.471 & 7.854 \\ \hline
  MOICBLS & 3.456 & 3.444 & 7.663 & 6.923 \\ \hline
  MG~DFR & 2.923 & 1.951 &\underline{3.927} & \underline{2.898} \\ \hline
  identity~DFR & \underline{2.898} & \underline{1.946} & 3.964 & 2.905 \\ \hline
  tanh~DFR & \bf{2.818} & \bf{1.930} & \bf{3.914} & \bf{2.886} \\ \hline
 \end{tabular}
\end{table}

\begin{figure}[tb]
 \centering
 \includegraphics[width=0.8\linewidth]{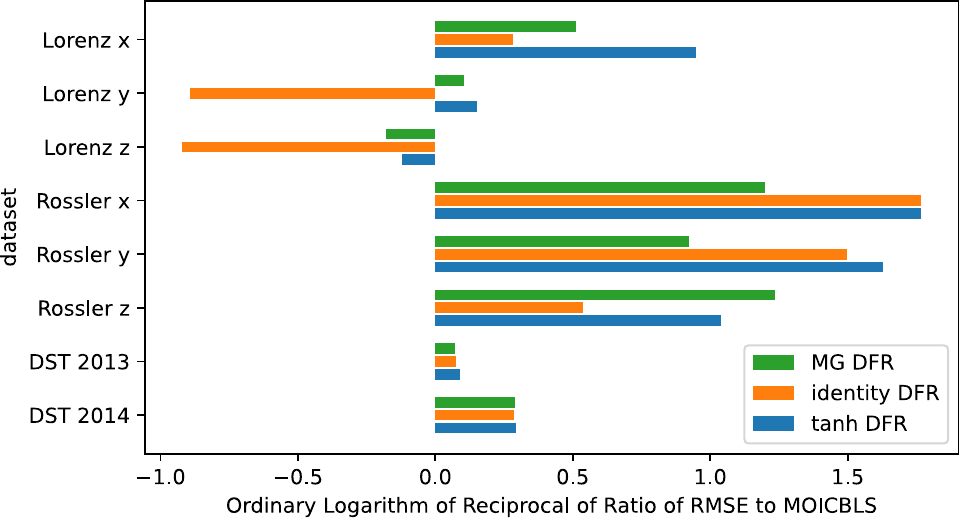}
 \caption{
 Logarithm of reciprocal of ratio of RMSE of DFR method to that of MOICBLS\@.
 A bar extending further to the right represents a superior method to MOICBLS,
 whereas positive or negative values indicate whether a method is better or worse, respectively, than MOICBLS\@.
 }
 \label{fig:ratio}
\end{figure}

Figure~\ref{fig:ratio} plots the logarithm of the reciprocal of the ratio of the RMSE of each DFR methods
to those of MOICBLS, the state-of-the-art method for all above datasets.
It shows that the further the bar extends to the right, the better is the method.
In the figure, the proposed tanh~DFR exhibits excellent time-series forecasting accuracy even more than the MG~DFR\@.
We also obeserve that the identity~DFR is as accurate as an MG~DFR and a tanh~DFR for most datasets.
For some datasets, such as Lorenz~$y$ and Lorenz~$z$, nonlinearity of $f$ is considered necessary to achieve high accuracy.
For future comparisons, evaluation results on additional datasets are presented in the Appendix.

\subsection{Comparison with existing hardware implementations}
The hardware implementation of the proposed identity~DFR model, which performs well on most datasets
and is the easier of the two proposed DFRs to implement, was evaluated.

\begin{figure}[tb]
 \centering
 \includegraphics[width=0.65\linewidth]{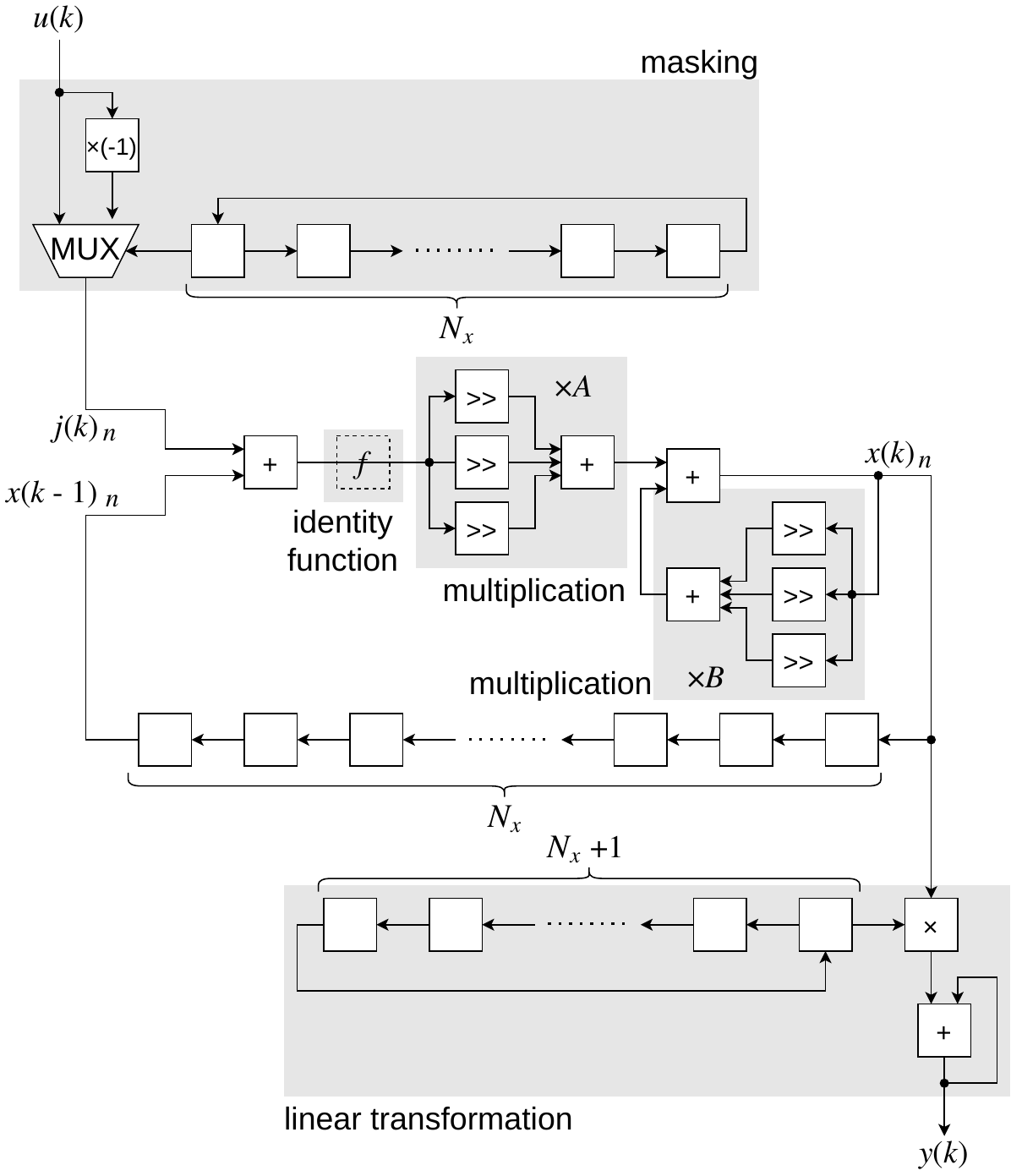}
 \caption{Implementation of the proposed identity~DFR\@.
 Blank blocks with cyclic arrow connection form a shift register.
 Shift for each approximation multiplier block is optimized.
 }\label{fig:propose}
\end{figure}

The proposed identity~DFR model (Prop.) was implemented in an FPGA board using Zynq-7000 (xc7z020clg400-1).
The detailed implementation of the identity~DFR is illustrated in Fig.~\ref{fig:propose}.
The masking procedure was implemented with a sign inverter, a multiplexer, and shift registers that store 1-bit weight.
The mask signal adopted two values, $+1$ and $-1$.
The mask processing was applied to one input signal $u(k)$, resulting in $N_x$ reservoir input signals $j(k)_n$.
In the following experiments, the number of virtual nodes in the reservoir, $N_x$, was set as 100.
An approximate multiplication using the sum of three right shifts with different shifts was adopted for acceleration and resource usage-saving.
$A$ and $B$ in Fig.~\ref{fig:propose} correspond to $\eta(1-\e^{-\theta})$ and $\e^{-\theta}$ in Fig.~\ref{fig:digitalDFR}, respectively.
The feedback in the reservoir was configured by the shift registers.
Here, for each input signal $u(k)$, $N_x$ reservoir features $x(k)_n$ were obtained.
Linear transformation was implemented with a multiplier, an adder, and shift registers storing the weight.
Each $x(k)_n$ was multiplied by the corresponding weight, $w_n$, and summed, followed by addition of a constant term to obtain the output, $y(k)$.

Fixed-point number representation was used throughout the experiments.
Different bit widths were used for the reservoir and output layers.
When the number of bits in the reservoir layer was $r$, the input, $u(k)$, was normalized;
therefore, its maximum value was $l=2^{r-2}$.
Similarly, when the number of bits in the output layer was $o$, the target output, $d(k)$, was normalized such that its maximum value was $m=2^{o-1}$.
The weights of the output layer were learned offline using Python.
The C++ code was synthesized in Xilinx Vitis HLS 2021.1, and the circuit was implemented and evaluated in Xilinx Vivado 2021.1.

The proposed identity~DFR was compared with state-of-the-art analog and digital implementations.
The analog implementation compared was a hybrid FPGA--ASIC DFR (hybrid~DFR)~\cite{shears2021hybrid}.
This hybrid~DFR employed a nonlinear element based on the MG model, which was realized as an analog circuit on a 180\,nm ASIC.
The remaining circuits, including the ADC, were implemented using the those on the same FPGA as in the proposed method,
whereas a discrete DAC component was used.
In the following evaluations, we adopted the same two tasks as in~\cite{shears2021hybrid}, namely, NARMA10 and spectrum sensing.
The algorithm of the proposed identity DFR is different from those of MG-based DFRs (MG~DFR and hybrid~DFR),
and simplifications such as approximate arithmetic operation and bitwidth reduction were employed in the identity~DFR.
Therefore, the accuracy of these tasks should be different from that when using the compared methods.

The digital implementation compared was~\cite{appeltant2011information} (naive~DFR), 
in which a DFR using the MG model is implemented using a 32-bit floating point number representation.
This corresponds to the implementation discussed in Section~\ref{subsec:digitalDFR}.
The input and output layers are the same as those used in the proposed method, with $N_x$=100.
None of the above implementations use pipelining.

In the naive~DFR, fixed parameters of $p=1$ and $\theta=0.2$ were employed for both datasets.
For the identity~DFR, multiplication by $B$ was approximated using the summation of the three right shifts, $\{1,2,4\}$.
Other parameters were optimized by grid search. The values of the parameters used are summarized in Table~\ref{tab:param2}.
Note that the indicated values are applicable before normalization of the input, $u(k)$.
For the identity~DFR, $\beta$ must be multiplied by $(l/\mathrm{max}_k{u(k)})^2$ based on Eq.~(\ref{eq:square}).
\begin{table}
 \centering
 \caption{Parameters of DFRs used for the comparison of hardware implementations.
 The $A$ value represents the amount of shift in the three right shifts that constitute multiplication by $A$.}\label{tab:param2}
 \begin{tabular}{c|ccc|cc} \hline
  dataset & \multicolumn{3}{c|}{Naive~DFR} & \multicolumn{2}{c}{Prop.}\\
  & $\gamma$ & $\eta$ & $\beta$ & $A$ & $\beta$ \\ \hline\hline
  NARMA10 & 0.05 & 0.5 & 1E-16 & \{4,5,11\} & 4E-12 \\ \hline
  Spectrum Sensing & 0.01 & 0.01 & 1E-3 & \{7,9,11\} & 1E+4 \\ \hline
 \end{tabular}
\end{table}

The results of the accuracy evaluation on NARMA10 are summarized in Table~\ref{tab:NARMA10}.
Here, the accuracy is compared in terms of normalized RMSE (NRMSE), as in~\cite{shears2021hybrid}.
\begin{equation}
 \mathrm{NRMSE}=\frac{||\boldsymbol{y}-\boldsymbol{d}||}{||\boldsymbol{y}||},
\end{equation}
where $\boldsymbol{y}$ is the vector of all outputs in the test data, and
$\boldsymbol{d}$ is the vector of the target outputs.
$||\cdot||$ is the Euclidean norm.
\begin{table}
 \centering
 \caption{Bit width configurations and accuracy comparison on the NARMA10 task.
 The NRMSE of hybrid~DFR is obtained from~\cite{shears2021hybrid}.}\label{tab:NARMA10}
 \begin{tabular}{c||c|c|c|c} \hline
  Method    & Naive~DFR  & Hybrid~DFR & Prop. (16-bit) & Prop. (24-bit) \\\hline
  Nonlinear & Mackey- & Mackey- & Not used & Not used \\
  Element & Glass & Glass & (Identity) & (Identity) \\\hline
  Number rep. in & 32\,bit & 16\,bit & 16\,bit & 24\,bit\\
  the reservoir & floating & fixed & fixed & fixed\\\hline
  Number rep. in & 32\,bit & 32\,bit & 21\,bit & 29\,bit\\
  the output layer & floating & fixed & fixed & fixed\\\hline
  NRMSE     & 0.14 & 0.21 & 0.20 & 0.14\\ \hline
 \end{tabular}
\end{table}
The proposed identity~DFR achieved high accuracy, even when nonlinear elements were substituted with the identity function.
When the 24-bit fixed-point representation was used,
the accuracy of the identity DFR was comparable to that of the naive DFR using the standard floating-point number representation.

We then compared the accuracy using spectrum sensing (Table~\ref{tab:spectrum}).
Here, the area under the curve (AUC) was evaluated.
\begin{table}
 \centering
 \caption{AUC comparison (spectrum sensing).
 The values of the hybrid~DFR are obtained from~\cite{shears2021hybrid}.
 Bold font indicates the best result.}\label{tab:spectrum}
 \begin{tabular}{c|c|c|c|c} \hline
  SNR (dB) & Antennae count & Naive~DFR & Hybrid~DFR & Prop. (16-bit) \\\hline\hline
  $-10$ & 2 & \textbf{0.920} &         0.913  & \textbf{0.920}\\ \cline{2-5}
  $-10$ & 4 & \textbf{0.995} &         0.994  & \textbf{0.995}\\ \cline{2-5}
  $-10$ & 6 & \textbf{1.000} &         0.999  & \textbf{1.000}\\ \hline
  $-15$ & 2 &         0.775  &         0.770  & \textbf{0.776}\\ \cline{2-5}
  $-15$ & 4 &         0.937  &         0.934  & \textbf{0.937}\\ \cline{2-5}
  $-15$ & 6 & \textbf{0.989} &         0.988  & \textbf{0.989}\\ \hline
  $-20$ & 2 & \textbf{0.630} &         0.603  & \textbf{0.630}\\ \cline{2-5}
  $-20$ & 4 & \textbf{0.764} & \textbf{0.764} & \textbf{0.764}\\ \cline{2-5}
  $-20$ & 6 & \textbf{0.871} & \textbf{0.871} & \textbf{0.871}\\ \hline
 \end{tabular}
\end{table}
The proposed identity~DFR (16-bit) achieved equal or better accuracy than the existing implementations
under all conditions in this task. 
When the reservoir used a 16-bit precision, the accuracy was equivalent to a software implementation using floating point numbers.

Subsequently, we evaluated the circuit size, power consumption, and throughput.
Owing to space limitations, only the evaluation results on NARMA10 are presented.
In these comparisons, the operation frequency has a deterministic impact.
The reported operation frequency of the hybrid~DFR is 10\,MHz~\cite{shears2021hybrid},
probably owing to the use of a DAC and an ADC\@.
Higher frequencies correspondingly increase the power consumption of the DAC.
Conversely, fully digital implementations, namely, the naive and identity~DFRs,
operate at significantly higher frequencies, such as 100\,MHz or higher.
Consequently, we compared these performances separately with those of existing analog and digital implementations.

\subsubsection{Comparison with analog implementation}\label{subsec:comp_analog}

Table~\ref{tab:circuit1} compares the circuit size and power consumption with those of the hybrid~DFR.
Here, the number of logic elements and power consumption for the digital circuits in the DFR, excluding the DAC, analog circuit, and ADC, 
was calculated by implementing each method on the same FPGA platform at an operating frequency of 10\,MHz.
\begin{table}[tb]
 \centering
 \caption{Comparison of circuit size and power consumption (NARMA10).
 Bit widths are the same as those in Table~\ref{tab:NARMA10}.}
 \label{tab:circuit1}
 \begin{tabular}{c|c|c} \hline
  No. of DFR & Hybrid DFR & Prop. (16-bit)\\ \hline\hline
  Slice LUTs & 270 & 312\\ \hline
  Slice Registers & 407 & 374 \\ \hline
  Slice & 134 & 126 \\ \hline
  Block RAM Tile & 2 & 1.5 \\ \hline
  DSPs & 3 & 1 \\ \hline
  Power in digital domain & 1\,mW & 1\,mW \\ \hline
 Power in analog domain & $\textgreater$ 10\,mW & -- \\ \hline
 \end{tabular}
\end{table}
Conversely, for the hybrid~DFR, the power consumption of the ADC, DAC, and ASIC were needed to be added to the power of the logic circuits.
The reported power consumption of the ASIC, which provides a nonlinear function, was 24.55\,\si{\micro\watt}~\cite{shears2021hybrid}.
The hybrid~DFR employed a 16-bit DAC with a reference voltage of 2.5\,V and a Zynq-7000 embedded 12-bit XADC with a reference voltage of 1\,V.
The maximum power consumption of the XADC was 36\,mW~\cite{XilinxXADC}.
A key metric to assess converter performance is the figure of merit (FoM)~\cite{murmann2008d},
defined by the following equation~\cite{murmann2008d}:
\begin{equation}
\mathrm{FoM}=\frac{P}{f_{\mathrm{s}}\cdot 2^{\mathrm{ENOB}}},
\end{equation}
where $P$ is the power, $f_{\mathrm{s}}$ is the Nyquist sampling rate, and
\begin{equation}
\mathrm{ENOB}=\frac{\mathrm{SNDR (dB)}-1.76}{6.02},
\end{equation}
where SNDR is the signal-to-noise-and-distortion ratio.
We estimated that the power of this ADC alone was 10\,mW\footnote{
The lowest limit of the FoMs for ADCs reported in~\cite{murmann2008d} was 100\,fJ/(conv-step).
Considering the specifications of the XADC (sampling frequency $f_{\mathrm{s}}$=10\,MHz and
effective number of bits ENOB=12), the lowest possible power consumption was 4\,mW when FoM=100\,fJ/(conv-step).
It was estimated to consume 10\,mW for a more realistic value of FoM=250\,fJ/(conv-step).}.
When considering the power usage of other components, including the discrete DAC, 
the power consumption of the proposed identity~DFR was at least 10 times smaller than that of the hybrid~DFR\@.

\subsubsection{Comparison with digital implementation}

Table~\ref{tab:circuit2} compares the circuit size and power consumption with those of a naive~DFR\@.
Both DFRs were implemented using the same FPGA and operated at a frequency of 100\,MHz.
\begin{table}[tb]
 \centering
 \caption{Comparison of circuit size and power consumption (NARMA10).
 Bit widths are the same as those in Table~\ref{tab:NARMA10}.}
 \label{tab:circuit2}
 \begin{tabular}{c|c|c} \hline
  No. of DFR & Naive DFR & Prop. (16-bit)\\ \hline\hline
  Slice LUTs & 5412 & 312\\ \hline
  Slice Registers & 7406& 374 \\ \hline
  Slice & 1983 & 131 \\ \hline
  Block RAM Tile & 3 & 1.5 \\ \hline
  DSPs & 27 & 1 \\ \hline
  Power & 58\,mW & 7\,mW \\ \hline
 \end{tabular}
\end{table}
According to the results,
the proposed identity~DFR is 8.3 times more power efficient than the naive DFR implementing the MG model.

\subsubsection{Comparison of the throughput}

Table~\ref{tab:throughput} presents the results of the throughput evaluation.
\begin{table}[tb]
 \centering
 \caption{Throughput comparison.
 Value of hybrid~DFR is obtained from~\cite{shears2021hybrid}.}
 \label{tab:throughput}
 \begin{tabular}{c|c|c|c} \hline
  Method & Hybrid DFR & Naive DFR & Prop. (16-bit)\\ \hline\hline
  Clock Frequency (MHz) & 10 & 100 & 100 \\ \hline
  Throughput (samples/s) & 1625 & 4670 & 8608\\ \hline
 \end{tabular}
\end{table}
Based on the table, the proposed identity~DFR achieves 5.3 and 1.8 times higher throughput than the hybrid implementation and naive~DFR, respectively.

\subsubsection{Discussion}\label{subsec:discussion}

This discussion focuses on  the effectiveness of the proposed identity~DFR
with respect to two specific tasks in~\cite{shears2021hybrid}.
Although the identity~DFR was proven be effective for these tasks, its accuracy slightly
lagged that of the DFR using the MG model for certain other tasks.
Therefore, for these tasks, it may be necessary to introduce nonlinearity in $f$.
For example, using the MG model as $f$ provides nonlinearity with tunable parameters.
However, even in such cases, the modular~DFR model proposed in Section~\ref{subsec:newmodel}
can provide a lighter implementation option
and facilitate a power-- and performance--efficient
design suitable for embedded processing.

\section{Conclusion}\label{sec:conclusion}

This paper proposed a new DFR model, called modular~DFR, suitable for efficient digital implementation.
This new model provides greater flexibility in selecting the nonlinear structure without increasing the complexity of the design.
As practical examples of the proposed DFR model, identity and tanh~DFRs were presented,
which use identity and hyperbolic tangent functions, respectively, as the pluggable nonlinear function of the proposed modular DFR model.
The software simulation results demonstrated that the proposed tanh~DFR outperforms
state-of-the-art chaotic time series prediction methods.
The identity~DFR also shows comparable accuracy to the tanh~DFR on most datasets.
Moreover, the implementation using an FPGA showed efficiency of the proposed identity~DFR.
It achieves 10x and 8.3x reduction in power consumption, and a 5.3x and 1.8x increase in throughput compared to
those of state-of-the-art analog and digital implementations, respectively.

\begin{acks}
This work was supported in part by the Japan Society for the Promotion of Science KAKENHI under Grant 23H03362.
\end{acks}

\bibliographystyle{ACM-Reference-Format}
\bibliography{ref}


\begin{thebibliography}{22}


\ifx \showCODEN    \undefined \def \showCODEN     #1{\unskip}     \fi
\ifx \showDOI      \undefined \def \showDOI       #1{#1}\fi
\ifx \showISBNx    \undefined \def \showISBNx     #1{\unskip}     \fi
\ifx \showISBNxiii \undefined \def \showISBNxiii  #1{\unskip}     \fi
\ifx \showISSN     \undefined \def \showISSN      #1{\unskip}     \fi
\ifx \showLCCN     \undefined \def \showLCCN      #1{\unskip}     \fi
\ifx \shownote     \undefined \def \shownote      #1{#1}          \fi
\ifx \showarticletitle \undefined \def \showarticletitle #1{#1}   \fi
\ifx \showURL      \undefined \def \showURL       {\relax}        \fi
\providecommand\bibfield[2]{#2}
\providecommand\bibinfo[2]{#2}
\providecommand\natexlab[1]{#1}
\providecommand\showeprint[2][]{arXiv:#2}

\bibitem[Alomar et~al\mbox{.}(2015)]%
        {alomar2015digital}
\bibfield{author}{\bibinfo{person}{Miquel~L Alomar}, \bibinfo{person}{Miguel~C
  Soriano}, \bibinfo{person}{Miguel Escalona-Mor{\'a}n},
  \bibinfo{person}{Vincent Canals}, \bibinfo{person}{Ingo Fischer},
  \bibinfo{person}{Claudio~R Mirasso}, {and} \bibinfo{person}{Jose~L
  Rossell{\'o}}.} \bibinfo{year}{2015}\natexlab{}.
\newblock \showarticletitle{Digital implementation of a single dynamical node
  reservoir computer}.
\newblock \bibinfo{journal}{\emph{IEEE Transactions on Circuits and Systems II:
  Express Briefs}} \bibinfo{volume}{62}, \bibinfo{number}{10}
  (\bibinfo{year}{2015}), \bibinfo{pages}{977--981}.
\newblock


\bibitem[Appeltant et~al\mbox{.}(2011)]%
        {appeltant2011information}
\bibfield{author}{\bibinfo{person}{Lennert Appeltant},
  \bibinfo{person}{Miguel~Cornelles Soriano}, \bibinfo{person}{Guy Van~der
  Sande}, \bibinfo{person}{Jan Danckaert}, \bibinfo{person}{Serge Massar},
  \bibinfo{person}{Joni Dambre}, \bibinfo{person}{Benjamin Schrauwen},
  \bibinfo{person}{Claudio~R Mirasso}, {and} \bibinfo{person}{Ingo Fischer}.}
  \bibinfo{year}{2011}\natexlab{}.
\newblock \showarticletitle{Information processing using a single dynamical
  node as complex system}.
\newblock \bibinfo{journal}{\emph{Nature Communications}} \bibinfo{volume}{2},
  \bibinfo{number}{1} (\bibinfo{year}{2011}), \bibinfo{pages}{1--6}.
\newblock


\bibitem[Appeltant et~al\mbox{.}(2014)]%
        {appeltant2014constructing}
\bibfield{author}{\bibinfo{person}{Lennert Appeltant}, \bibinfo{person}{Guy
  Van~der Sande}, \bibinfo{person}{Jan Danckaert}, {and} \bibinfo{person}{Ingo
  Fischer}.} \bibinfo{year}{2014}\natexlab{}.
\newblock \showarticletitle{Constructing optimized binary masks for reservoir
  computing with delay systems}.
\newblock \bibinfo{journal}{\emph{Scientific Reports}} \bibinfo{volume}{4},
  \bibinfo{number}{1} (\bibinfo{year}{2014}), \bibinfo{pages}{1--5}.
\newblock


\bibitem[Banerjee et~al\mbox{.}(2012)]%
        {banerjee2012existence}
\bibfield{author}{\bibinfo{person}{Adrija Banerjee}, \bibinfo{person}{Amaresh
  Bej}, {and} \bibinfo{person}{TN Chatterjee}.}
  \bibinfo{year}{2012}\natexlab{}.
\newblock \showarticletitle{On the existence of a long range correlation in the
  geomagnetic disturbance storm time (Dst) index}.
\newblock \bibinfo{journal}{\emph{Astrophysics and Space Science}}
  \bibinfo{volume}{337} (\bibinfo{year}{2012}), \bibinfo{pages}{23--32}.
\newblock


\bibitem[Chandra(2015)]%
        {chandra2015competition}
\bibfield{author}{\bibinfo{person}{Rohitash Chandra}.}
  \bibinfo{year}{2015}\natexlab{}.
\newblock \showarticletitle{Competition and collaboration in cooperative
  coevolution of Elman recurrent neural networks for time-series prediction}.
\newblock \bibinfo{journal}{\emph{IEEE Transactions on Neural Networks and
  Learning Systems}} \bibinfo{volume}{26}, \bibinfo{number}{12}
  (\bibinfo{year}{2015}), \bibinfo{pages}{3123--3136}.
\newblock


\bibitem[Chen and Liu(2017)]%
        {chen2017broad}
\bibfield{author}{\bibinfo{person}{CL~Philip Chen} {and}
  \bibinfo{person}{Zhulin Liu}.} \bibinfo{year}{2017}\natexlab{}.
\newblock \showarticletitle{Broad learning system: An effective and efficient
  incremental learning system without the need for deep architecture}.
\newblock \bibinfo{journal}{\emph{IEEE Transactions on Neural Networks and
  Learning Systems}} \bibinfo{volume}{29}, \bibinfo{number}{1}
  (\bibinfo{year}{2017}), \bibinfo{pages}{10--24}.
\newblock


\bibitem[Chen and Lee(2015)]%
        {chen2015weighted}
\bibfield{author}{\bibinfo{person}{Thao-Tsen Chen} {and}
  \bibinfo{person}{Shie-Jue Lee}.} \bibinfo{year}{2015}\natexlab{}.
\newblock \showarticletitle{A weighted LS-SVM based learning system for time
  series forecasting}.
\newblock \bibinfo{journal}{\emph{Information Sciences}}  \bibinfo{volume}{299}
  (\bibinfo{year}{2015}), \bibinfo{pages}{99--116}.
\newblock


\bibitem[for Geomagnetism and Magnetism(2022)]%
        {DST}
\bibfield{author}{\bibinfo{person}{Data Analysis~Center for Geomagnetism} {and}
  \bibinfo{person}{Space Magnetism}.} \bibinfo{year}{2022}\natexlab{}.
\newblock \bibinfo{title}{Plot and data output of Dst and AE indices (Hourly
  Values)}.
\newblock
\newblock
\newblock
\shownote{\url{https://wdc.kugi.kyoto-u.ac.jp/dstae/index.html}}.


\bibitem[Guo et~al\mbox{.}(2019)]%
        {guo2019parallel}
\bibfield{author}{\bibinfo{person}{Pei Guo}, \bibinfo{person}{Chen Liuy},
  \bibinfo{person}{Yan Tang}, {and} \bibinfo{person}{Jianwu Wang}.}
  \bibinfo{year}{2019}\natexlab{}.
\newblock \showarticletitle{Parallel gradient boosting based granger causality
  learning}. In \bibinfo{booktitle}{\emph{IEEE International Conference on Big
  Data}}. IEEE, \bibinfo{pages}{2845--2854}.
\newblock


\bibitem[Han and Xu(2017)]%
        {han2017laplacian}
\bibfield{author}{\bibinfo{person}{Min Han} {and} \bibinfo{person}{Meiling
  Xu}.} \bibinfo{year}{2017}\natexlab{}.
\newblock \showarticletitle{Laplacian echo state network for multivariate time
  series prediction}.
\newblock \bibinfo{journal}{\emph{IEEE Transactions on Neural Networks and
  Learning Systems}} \bibinfo{volume}{29}, \bibinfo{number}{1}
  (\bibinfo{year}{2017}), \bibinfo{pages}{238--244}.
\newblock


\bibitem[Ikeda et~al\mbox{.}(2022)]%
        {ikeda2022hardware}
\bibfield{author}{\bibinfo{person}{Sosei Ikeda}, \bibinfo{person}{Hiromitsu
  Awano}, {and} \bibinfo{person}{Takashi Sato}.}
  \bibinfo{year}{2022}\natexlab{}.
\newblock \showarticletitle{Hardware-friendly delayed-feedback reservoir for
  multivariate time-series classification}.
\newblock \bibinfo{journal}{\emph{IEEE Transactions on Computer-Aided Design of
  Integrated Circuits and Systems}} \bibinfo{volume}{41}, \bibinfo{number}{11}
  (\bibinfo{year}{2022}), \bibinfo{pages}{3650--3660}.
\newblock
\urldef\tempurl%
\url{https://doi.org/10.1109/TCAD.2022.3197488}
\showDOI{\tempurl}


\bibitem[Laumanns et~al\mbox{.}(2002)]%
        {laumanns2002combining}
\bibfield{author}{\bibinfo{person}{Marco Laumanns}, \bibinfo{person}{Lothar
  Thiele}, \bibinfo{person}{Kalyanmoy Deb}, {and} \bibinfo{person}{Eckart
  Zitzler}.} \bibinfo{year}{2002}\natexlab{}.
\newblock \showarticletitle{Combining convergence and diversity in evolutionary
  multiobjective optimization}.
\newblock \bibinfo{journal}{\emph{Evolutionary Computation}}
  \bibinfo{volume}{10}, \bibinfo{number}{3} (\bibinfo{year}{2002}),
  \bibinfo{pages}{263--282}.
\newblock


\bibitem[Luko{\v{s}}evi{\v{c}}ius and Jaeger(2009)]%
        {lukovsevivcius2009reservoir}
\bibfield{author}{\bibinfo{person}{Mantas Luko{\v{s}}evi{\v{c}}ius} {and}
  \bibinfo{person}{Herbert Jaeger}.} \bibinfo{year}{2009}\natexlab{}.
\newblock \showarticletitle{Reservoir computing approaches to recurrent neural
  network training}.
\newblock \bibinfo{journal}{\emph{Computer Science Review}}
  \bibinfo{volume}{3}, \bibinfo{number}{3} (\bibinfo{year}{2009}),
  \bibinfo{pages}{127--149}.
\newblock


\bibitem[Mackey and Glass(1977)]%
        {mackey1977oscillation}
\bibfield{author}{\bibinfo{person}{Michael~C Mackey} {and}
  \bibinfo{person}{Leon Glass}.} \bibinfo{year}{1977}\natexlab{}.
\newblock \showarticletitle{Oscillation and chaos in physiological control
  systems}.
\newblock \bibinfo{journal}{\emph{Science}} \bibinfo{volume}{197},
  \bibinfo{number}{4300} (\bibinfo{year}{1977}), \bibinfo{pages}{287--289}.
\newblock


\bibitem[Murmann(2008)]%
        {murmann2008d}
\bibfield{author}{\bibinfo{person}{Boris Murmann}.}
  \bibinfo{year}{2008}\natexlab{}.
\newblock \showarticletitle{{A/D} converter trends: Power dissipation, scaling
  and digitally assisted architectures}. In \bibinfo{booktitle}{\emph{IEEE
  Custom Integrated Circuits Conference}}. \bibinfo{pages}{105--112}.
\newblock


\bibitem[Shears et~al\mbox{.}(2021)]%
        {shears2021hybrid}
\bibfield{author}{\bibinfo{person}{Osaze Shears}, \bibinfo{person}{Kangjun
  Bai}, \bibinfo{person}{Lingjia Liu}, {and} \bibinfo{person}{Yang Yi}.}
  \bibinfo{year}{2021}\natexlab{}.
\newblock \showarticletitle{A Hybrid {FPGA-ASIC} delayed feedback reservoir
  system to enable spectrum sensing/sharing for low power {IoT} devices}. In
  \bibinfo{booktitle}{\emph{IEEE/ACM International Conference on Computer Aided
  Design}}. \bibinfo{pages}{1--9}.
\newblock


\bibitem[Soriano et~al\mbox{.}(2014)]%
        {soriano2014delay}
\bibfield{author}{\bibinfo{person}{Miguel~C Soriano}, \bibinfo{person}{Silvia
  Ort{\'\i}n}, \bibinfo{person}{Lars Keuninckx}, \bibinfo{person}{Lennert
  Appeltant}, \bibinfo{person}{Jan Danckaert}, \bibinfo{person}{Luis Pesquera},
  {and} \bibinfo{person}{Guy Van~der Sande}.} \bibinfo{year}{2014}\natexlab{}.
\newblock \showarticletitle{Delay-based reservoir computing: Noise effects in a
  combined analog and digital implementation}.
\newblock \bibinfo{journal}{\emph{IEEE Transactions on Neural Networks and
  Learning Systems}} \bibinfo{volume}{26}, \bibinfo{number}{2}
  (\bibinfo{year}{2014}), \bibinfo{pages}{388--393}.
\newblock


\bibitem[Tanaka et~al\mbox{.}(2019)]%
        {tanaka2019recent}
\bibfield{author}{\bibinfo{person}{Gouhei Tanaka}, \bibinfo{person}{Toshiyuki
  Yamane}, \bibinfo{person}{Jean~Benoit H{\'e}roux}, \bibinfo{person}{Ryosho
  Nakane}, \bibinfo{person}{Naoki Kanazawa}, \bibinfo{person}{Seiji Takeda},
  \bibinfo{person}{Hidetoshi Numata}, \bibinfo{person}{Daiju Nakano}, {and}
  \bibinfo{person}{Akira Hirose}.} \bibinfo{year}{2019}\natexlab{}.
\newblock \showarticletitle{Recent advances in physical reservoir computing: A
  review}.
\newblock \bibinfo{journal}{\emph{Neural Networks}}  \bibinfo{volume}{115}
  (\bibinfo{year}{2019}), \bibinfo{pages}{100--123}.
\newblock


\bibitem[Xilinx(2022)]%
        {XilinxXADC}
\bibfield{author}{\bibinfo{person}{Xilinx}.} \bibinfo{year}{2022}\natexlab{}.
\newblock \bibinfo{title}{7 Series {FPGAs} and {Zynq-7000} {SoC} {XADC} {Dual}
  12-Bit 1 {MSPS} Analog-to-Digital Converter User Guide ({UG480})}.
\newblock
\newblock
\newblock
\shownote{\url{https://docs.xilinx.com/r/en-US/ug480_7Series_XADC}}.


\bibitem[Xue et~al\mbox{.}(2019)]%
        {xue2019semg}
\bibfield{author}{\bibinfo{person}{Yaxu Xue}, \bibinfo{person}{Xiaofei Ji},
  \bibinfo{person}{Dalin Zhou}, \bibinfo{person}{Jing Li}, {and}
  \bibinfo{person}{Zhaojie Ju}.} \bibinfo{year}{2019}\natexlab{}.
\newblock \showarticletitle{SEMG-based human in-hand motion recognition using
  nonlinear time series analysis and random forest}.
\newblock \bibinfo{journal}{\emph{IEEE Access}}  \bibinfo{volume}{7}
  (\bibinfo{year}{2019}), \bibinfo{pages}{176448--176457}.
\newblock


\bibitem[Yan et~al\mbox{.}(2017)]%
        {yan2017fast}
\bibfield{author}{\bibinfo{person}{Ke Yan}, \bibinfo{person}{Zhiwei Ji},
  \bibinfo{person}{Huijuan Lu}, \bibinfo{person}{Jing Huang},
  \bibinfo{person}{Wen Shen}, {and} \bibinfo{person}{Yu Xue}.}
  \bibinfo{year}{2017}\natexlab{}.
\newblock \showarticletitle{Fast and accurate classification of time series
  data using extended ELM: Application in fault diagnosis of air handling
  units}.
\newblock \bibinfo{journal}{\emph{IEEE Transactions on Systems, Man, and
  Cybernetics: Systems}} \bibinfo{volume}{49}, \bibinfo{number}{7}
  (\bibinfo{year}{2017}), \bibinfo{pages}{1349--1356}.
\newblock


\bibitem[Yi et~al\mbox{.}(2022)]%
        {yi2022intergroup}
\bibfield{author}{\bibinfo{person}{Jun Yi}, \bibinfo{person}{Jiahua Huang},
  \bibinfo{person}{Wei Zhou}, \bibinfo{person}{Guorong Chen}, {and}
  \bibinfo{person}{Meng Zhao}.} \bibinfo{year}{2022}\natexlab{}.
\newblock \showarticletitle{Intergroup cascade broad learning system with
  optimized parameters for chaotic time series prediction}.
\newblock \bibinfo{journal}{\emph{IEEE Transactions on Artificial
  Intelligence}} \bibinfo{volume}{3}, \bibinfo{number}{5}
  (\bibinfo{year}{2022}), \bibinfo{pages}{709--721}.
\newblock
\urldef\tempurl%
\url{https://doi.org/10.1109/TAI.2022.3143079}
\showDOI{\tempurl}


\end{thebibliography}

\appendix

\section{Additional evaluation for future comparisons}
This section presents the evaluation results on additional datasets for future comparisons.

The datasets used comprised the most recent 2021 and 2022 DST datasets, along with the Chen system, a typical chaotic system.
For the DST datasets, 4,000 samples were selected from the tails of both the 2021 and 2022 data.
The initial 3,000 samples were employed for training,
whereas the remaining 1,000 samples were reserved for testing, following the methodology described in Section~\ref{sec:evaluation}.

The Chen system is defined as:
\begin{eqnarray}
 \frac{{\mathrm d}x}{{\mathrm d}t}&=&a(y-x),\nonumber \\
 \frac{{\mathrm d}y}{{\mathrm d}t}&=&(c-a)x-xz+cy,\\
 \frac{{\mathrm d}z}{{\mathrm d}t}&=&xy-bz,\nonumber
\end{eqnarray}
where $a=40$, $b=3$, $c=28$, $x(0)=-0.1$, $y(0)=0.5$, and $z(0)=-0.6$ in the following experiments.
The fourth-order Runge--Kutta method was used to generate 12,000 samples with a sampling interval of 0.01~s.
Among these, the first 9,000 samples were used for training, and the remaining 3,000 samples were used for testing.
The DFRs were trained to predict the value of $x$.

The evaluation conditions were the same as in Section~\ref{sec:evaluation}, and
the parameter values used are summarized in Table~\ref{tab:param_appendix}.
\begin{table}
 \centering
 \caption{Parameters of DFRs used for additional datasets.}
 \label{tab:param_appendix}
 \begin{tabular}{c|ccc|cc|ccc} \hline
  dataset & \multicolumn{3}{c|}{MG~DFR} & \multicolumn{2}{c|}{identity~DFR} & \multicolumn{3}{c}{tanh~DFR}\\
  & $\gamma$ & $\eta$ & $\beta$ & $A$ & $\beta$ & $A$ & $\alpha$ & $\beta$ \\ \hline\hline
  DST~2021 & 4E-4 & 0.4 & 1E-18 & 0.004 & 1E-6 & 0.03 & 2 & 1E-5 \\ \hline
  DST~2022 & 5E-4 & 0.4 & 1E-18 & 0.002 & 1E-5 & 0.09 & 2 & 1E-5 \\ \hline
  Chen\_x & 0.03 & 0.3 & 1E-12 & 0.03 & 1E-7 & 0.07 & 7 & 1E-5 \\ \hline
 \end{tabular}
\end{table}

The accuracy evaluation results are summarized in Table~\ref{tab:result_appendix}.
\begin{table}
 \centering
 \caption{Prediction results of DFRs on additional datasets.}
 \label{tab:result_appendix}
 \begin{tabular}{c|cc|cc|cc} \hline
  & \multicolumn{2}{c|}{DST~2021} & \multicolumn{2}{c|}{DST~2022} & \multicolumn{2}{c}{Chen\_x}\\
  methods & RMSE & MAE & RMSE & MAE & RMSE & MAE \\ \hline\hline
  MG~DFR & 3.141 & 2.116 & 3.782 & 2.784 & 1.259E-4 & 9.169E-5 \\ \hline
  identity~DFR & 3.171 & 2.123 & 3.841 & 2.831 & 7.851E-4 & 4.783E-4 \\ \hline
  tanh~DFR & 3.146 & 2.117 & 3.792 & 2.773 & 1.156E-4 & 6.158E-5 \\ \hline
 \end{tabular}
\end{table}

\begin{figure}[tb]
 \centering
 \includegraphics[width=0.8\linewidth]{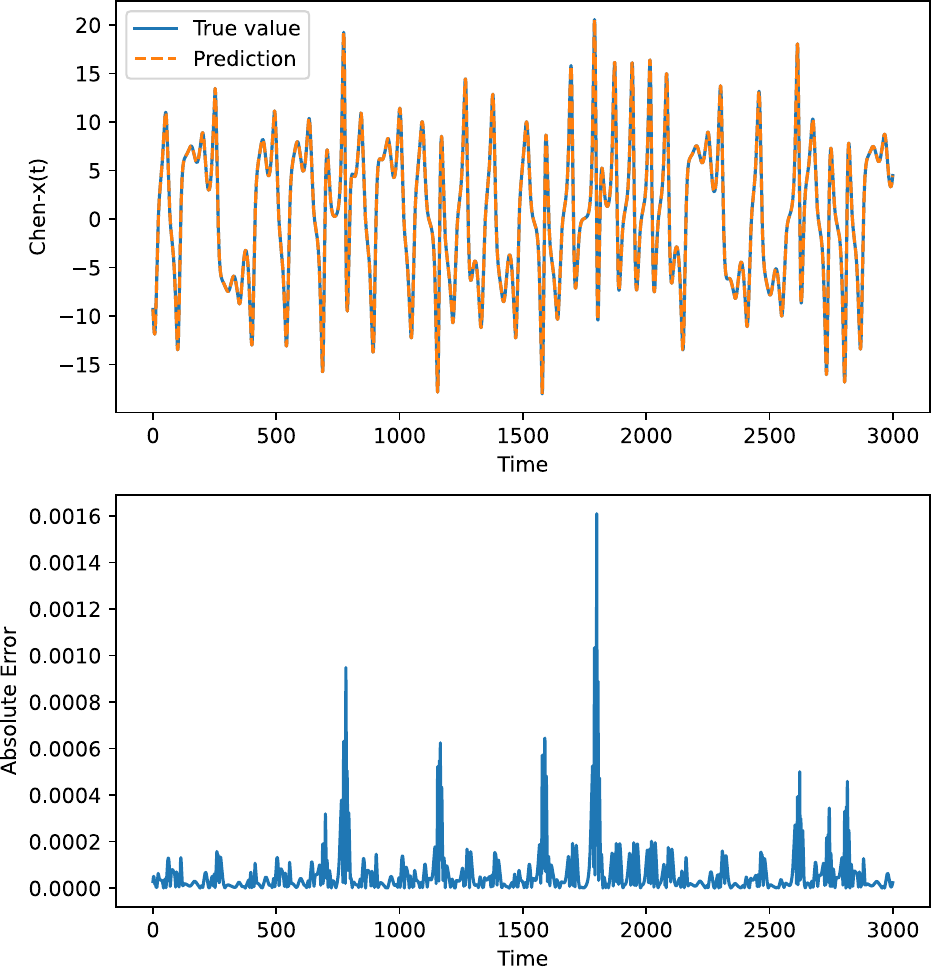}
 \caption{Comparison of true and predicted values and absolute error by tanh~DFR for the Chen system.}
 \label{fig:Chen}
\end{figure}

The forecast result for the Chen system using the proposed tanh~DFR is shown in Figure~\ref{fig:Chen}.
The upper figure compares the predicted and true values, and the lower figure plots the absolute error.

\end{document}